\title{Higher Order Correlations within Cortical Layers Dominate Functional Connectivity in Microcolumns}
\author{
Urs K\"oster \\ 
Redwood Center for Theoretical Neuroscience \\
UC Berkeley 
\And 
Jascha Sohl-Dickstein \\ 
Department of Applied Physics \\
Stanford University
\And 
Charles M Gray \\ 
Cellular Biology and Neuroscience\\ 
Montana State University Bozeman
\And 
Bruno A Olshausen \\ 
Redwood Center for Theoretical Neuroscience \\
UC Berkeley
}
 
\date{\today}

\documentclass[10pt]{article} % For LaTeX2e
\usepackage{nips12submit_e,times} % anonymized submission style files
\usepackage{amsmath,amssymb}
\usepackage{graphicx}
\usepackage{color}
\definecolor{orange}{rgb}{1,0.5,0}
\definecolor{gray}{rgb}{0.5,0.5,0.5}
\definecolor{blue}{rgb}{0.0,0.0,1.0}

\newcommand{\pd}[2]{\frac{\partial #1}{\partial #2}}
\newcommand{\mb}{\mathbf}

\usepackage{verbatim}

\begin{document}
\maketitle

\begin{abstract}
% Revision from Bruno
We report on simultaneous recordings from cells in all layers of visual cortex and models developed to capture the higher order structure of population spiking activity. Specifically, we use Ising, Restricted Boltzmann Machine (RBM) and semi-Restricted Boltzmann Machine (sRBM) models to reveal laminar patterns of activity.  While the Ising model describes only pairwise couplings, the RBM and sRBM capture higher-order dependencies using hidden units.
%Estimation of the models is made practical using Minimum Probability Flow, a recently developed parameter estimation method for energy-based models. The partition functions of the models are estimated using annealed importance sampling, which allows models to be compared in terms of likelihood. 
Applied to 32-channel polytrode data recorded from cat visual cortex, the higher-order models discover functional connectivity preferentially linking groups of cells within a cortical layer. % JSD what is a "layer group"?  maybe just "layers"?
Both RBM and sRBM models outperform Ising models in log-likelihood. Additionally, we train all three models on spatiotemporal sequences of states, exposing temporal structure and allowing us to predict spiking from network history. 
%Higher order interactions become far more important in the spatiotemporal case, and the models achieve far higher log likelihood as well as exposing temporal structure.  
This demonstrates the importance of modeling higher order interactions across space and time when characterizing activity in cortical networks. %and its relation to laminar architecture.

\end{abstract}

% ----------------------------------
% INTRODUCTION
% ----------------------------------

\section{Introduction}

% this is the brain...
Electrophysiology is rapidly moving towards high density recording techniques capable of capturing the simultaneous activity of large populations of neurons. This raises the challenge of understanding how networks encode and process information in ways that go beyond feedforward receptive field models for individual neurons. Modeling the distribution of states in a network provides a way to discover communication patterns and functional connectivity between cells. 

%prior art -- modify this to lead up to RBM but no GML and HMM here!
The Ising model, originally developed in the 1920s to describe magnetic interactions~\cite{Ising1925}, has been used to describe electrophysiological data, particularly in the retina~\cite{Schneidman2006}, and more recently for cortical recordings \cite{Yu2008}. This model treats spikes from a population of neurons binned in time as binary vectors and captures dependencies between cells with the maximum entropy distribution for pairwise correlation. However, this only provides a good approximation for small groups of cells in the retina~\cite{Ganmor2011}. % as it is difficult to extend to capture higher-order correlations.  (the reason it only applies to small groups isn't that it's diffucult to extend.  if you want to keep the seccond half of the sentence, split it into it its own sentence.

% [THE PROBLEM]
In this work, we apply maximum entropy models to data from the visual cortex. 
%Cortex is less well-understood than retina or LGN: 
Cortical networks have proven to be more challenging to model than the retina:
even the existence of significant pairwise correlations between cortical cells is controversial~\cite{Ecker2010,Renart2010} and  higher order correlations play an important role~\cite{Ohiorhenuan2010,Ohiorhenuan2011,Yu2011}. 
% sparse sampling
One of the challenges with current recording technologies is that we cannot simultaneously record from more than a tiny fraction of the cells that make up a cortical circuit. 
Sparse sampling together with the complexity of the circuit mean that the majority of a cell's input will be from cells that are not part of the recording. In adult cat visual cortex, direct synaptic connections have been reported to occur between 11\% - 30\% of nearby pairs of excitatory neurons in layer IV~\cite{Stratford1996}, while a larger fraction of cell pairs show ``polysynaptic'' couplings \cite{Ghose1994}, defined by a broad peak in the cross-correlation between two cells. This type of coupling can be due to common inputs (either from a different cortical area or in the form of lateral connections) or a series of monosynaptic connections. % JSD what is a lateral interaction? 
A combination of these is believed to give rise to most of the statistical interactions between the recorded cells. The Ising model, which assumes only pairwise couplings, is well suited to model direct synaptic coupling, but cannot deal with interactions that include many cells simultaneously. Therefore, we propose a new approach, that addresses both incomplete sampling and tentative inputs from larger scale cell assemblies. We extend the Ising model with a layer of hidden units or latent variables. 
The resulting model is a semi-Restricted Boltzmann Machine (sRBM), which combines pairwise connections between visible units with an additional set of connections to hidden units. %, as in a restricted Boltzmann machine (RBM,~\cite{Hinton1986}).

% [THE SOLUTION]
%[b] Learning is hard and 
Estimating the parameters of Ising models and Boltzmann machines is hard because %evaluating model probabilities involves computing a shared normalization constant. 
probabilities are only defined up to a normalization constant. 
For both Ising models and Boltzmann machines with hidden units, this normalization constant is computationally intractable, requiring a sum over the exponential number of states of the system. This makes exact maximum likelihood estimation impossible for all but the smallest systems and has previously necessitated approximate and/or computationally expensive estimation methods. % based on sampling or approximations of the true likelihood.
% MPF
In this work, we use Minimum Probability Flow (MPF~\cite{Sohl-Dickstein2011,Sohl-Dickstein2011b}, in the context of neural decoding see \cite{Schaub2012}) to estimate parameters efficiently without computing the intractable partition function. This allows us to estimate Ising models on higher-dimensional data than is otherwise feasible, and to estimate the sRBM in a straightforward way. 
% add a blurb about AIS
Similar to parameter estimation, model evaluation is also hampered by the fact that the learned models are not normalized. To compute probabilities and compare the likelihood of different models, annealed importance sampling~\cite{Neal2001} was used to estimate the partition function. 

% and conclude a tiny bit
Combining these two methods for model estimation and evaluation, we show that the sRBM can capture the distribution of states in a cortical network of tens of cells recorded from cat visual cortex significantly better than a pairwise model. The higher order structure discovered by the model is spatially organized and specific to cortical layers, indicating that networks within individual layers of a microcolumn are the dominant source of correlations. 
% JSD add some text here about the spatiotemporal version of the model?
Applied to spatiotemporal patterns of activity, the model captures temporal structure in addition to dependencies across different cells, allowing us to predict spiking activity based on the history of the network.

% ----------------------------------
% METHODS
% ----------------------------------
\section{Methods}

% a little bit about the experiment!!
\subsection{Recording and experimental procedures}

% fig1: show some data.
\begin{figure*}[t] 
  \centering
  \includegraphics{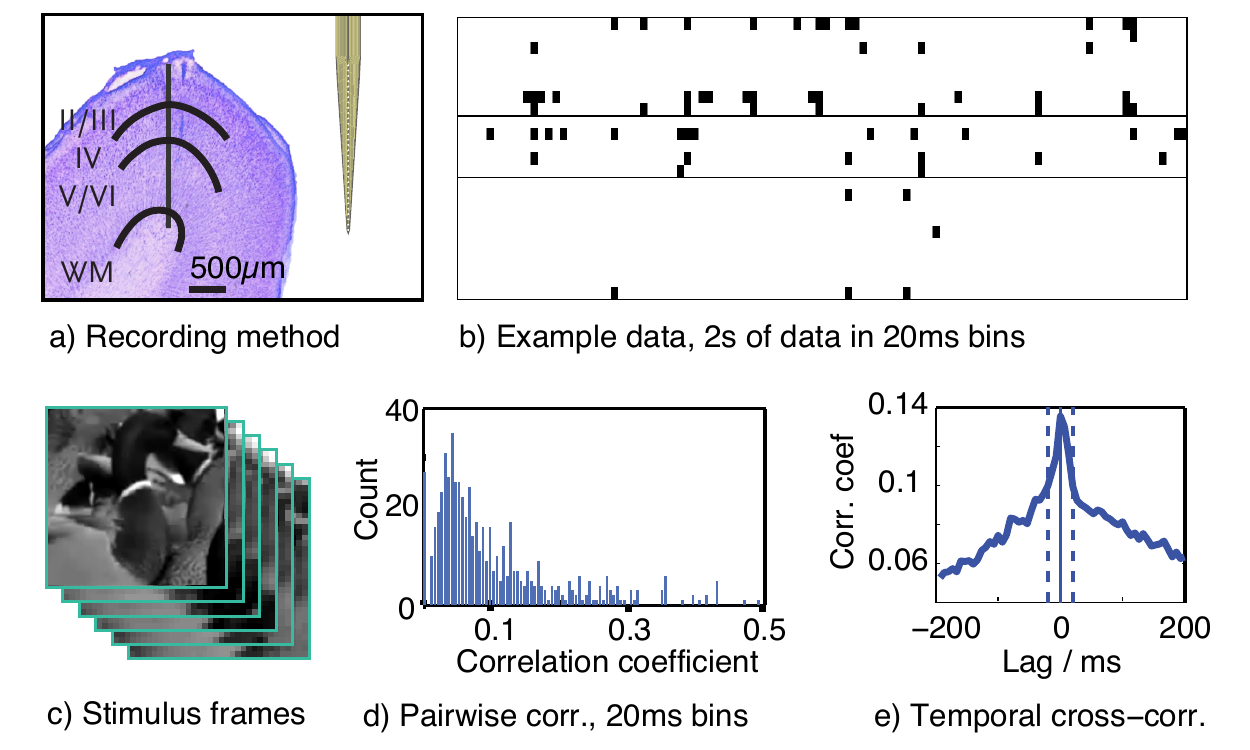} 
  \caption{ \textbf{(a)} Recordings were made from columns of cat visual cortex. The 32 channel probe is shown to scale. \textbf{(b)} Example data from 23 cells, binned in 20ms windows, 2s of data. Columns of this matrix are the input to our algorithm. For the spatiotemporal version of the model, we concatenate several adjacent columns.  % JSD add some text here about the spatiotemporal version of the model?
  \textbf{(c)} Example of one of the natural movie stimuli presented, showing a group of ducks. \textbf{(d)} Histogram of instantaneous correlations between pairs of cells. \textbf{(e)} Cross-correlation as a function of time lag between a pair of cells binned at 6.7ms shows a peak in the correlation about 20ms wide (dashed lines). This determines the window size for binning the data.  }
  \label{fig:data} 
\end{figure*}
% stimulus does not really matter here since it's not a response model, just mention it in one senence:
Data were recorded from anesthetized cat visual cortex in response to a custom set of full field natural movie stimuli. Movies of 8 to 30 minutes duration were captured at 300 frames per second and $512 \times 384$ pixel resolution with a Casio F1 camera. Care was taken to avoid scene changes and camera motion, so the spatiotemporal statistics of the movie correspond to those of natural scene motion. Movies were converted to grayscale and down-sampled to 150 Hz for presentation. The high frame rate was chosen to prevent cells phase locking to the frame rate, and scene changes were minimized to avoid evoked potentials due to sudden luminance changes. Fig \ref{fig:data}c) shows an example frame crop from one of the natural scene movies. 

% one sentence about recording methods
Recordings were made with single shank 32 channel polytrodes  (Fig. \ref{fig:data}a) with a channel spacing of 50$\mu$m, spanning all the layers of visual cortex. Individual datasets had on the order of of 20 to 40 simultaneously recorded neurons. The data was spike sorted offline using k-means clustering (KlustaKwik, \cite{Harris2000}) with a manual cleanup step (MClust, \cite{Redish}). Unless noted otherwise, spikes were binned at 20ms where bins with a single spike (2.4\% of bins) and multiple spikes (0.9\% of bins) were both treated as spiking and the rest as non-spiking. The bin size was chosen to span the width of the central peak in cross-correlograms between pairs of cells such as shown in \ref{fig:data}e), where the dashed vertical lines at $\pm$ 20ms envelope the central peak. An example of binned data for 23 isolated single units is shown in \ref{fig:data}b) with black marks indicating spiking.
%An example is shown in Fig \ref{fig:data}e) for a cell pair that has a strong instantaneous correlation with a central peak about 20ms wide.

% one sentence about spike sorting (def. methods!) CSD reconstruction etc. 
To register individual recording channels with cortical layers, recording locations were reconstructed from Nissl-stained histological sections, and current source density analysis in response to flashed full-field stimuli was used to infer the location of cortical layer IV on the polytrode \cite{Mitzdorf1985}. In \ref{fig:data}b) we use horizontal lines to indicate the upper and lower boundaries of layer IV. 

% include this??
The surgical methods are described in detail elsewhere \cite{Gray1995}. The protocol used in the experiments was approved by the Institutional Animal Care and Use Committee at Montana State University and conformed to the guidelines recommended in Preparation and Maintenance of Higher Mammals During Neuroscience Experiments, National Institutes of Health Publication 91–3207 (National Institutes of Heath, Bethesda, MD 1991).

% this paragraph briges over to modeling. 
The models were estimated on a data set of $180,000$ data vectors corresponding to 60 minutes of recording time. The data were split into two subsets of $90,000$ data points: a training set for parameter estimation and a test set to compute cross-validated likelihoods. 

We also analyzed spatiotemporal patterns of data, which were created by concatenating consecutive state vectors. For the spatiotemporal experiments a bin width of 6.7 ms, corresponding to the frame rate of the stimulus, was used. This bin size is a compromise capturing more detailed structure in the data without leading to an undue increase in dimensionality and complexity. %JSD why 6.7 ms?  (Is this so there's enough training data?)
Up to 10 time bins were concatenated in order to discover spatiotemporal patterns and predict spiking given the history over the prior 67ms. These models were trained on $100,000$ samples and likelihoods were computed on a set of equal size.

\subsection{Model and Estimation}

% sRBM
The sRBM consists of a set of binary visible units $\mb x \in \{ 0, 1 \}^N$ corresponding to observed neurons in the data and a set of hidden units $\mb h \in \{ 0, 1 \}^M$ that capture higher order dependencies. Weights between visible units, corresponding to an Ising model or fully visible Boltzmann machine, capture pairwise correlations in the data. Weights between visible and hidden units, corresponding to an RBM, learn to describe higher order structure. 

The Ising model with visible-visible coupling weights $\mb J \in \mathcal R^{N \times N}$ and biases $\mb b \in \mathcal R^N$ has an energy function 
% we flip the sign here to be consistent with Schneidmann, but really we just do imagesc(-J) and don't use this in estimation
\begin{equation}
    E_{\mathrm{I}}(\mb x) =  -\mb x^T \mb J \mb x - \mb b^T \mb  x 
   ,
\end{equation}
with associated probability distribution $p_{\mathrm{I}}\left( \mb x \right) = \frac{1}{Z_{\mathrm{I}}}\exp\left[  -E_{\mathrm{I}}(\mb x) \right]$, where the normalization constant, or partition function, $Z_{\mathrm{I}} = \sum_{\left\{ \mb x' \right\}} \exp\left[  -E_{\mathrm{I}}(\mb x') \right]$ consists of a sum over all $2^N$ system states.

The RBM with visible-hidden coupling weights $\mb W \in \mathcal R^{N \times M}$ and hidden and visible biases $\mb b_v \in \mathcal R^N$ and $\mb b_h \in \mathcal R^M$ has an energy function
\begin{equation}
    E_{\mathrm{R}}(\mb x, \mb h) = - \mb x^T \mb W \mb h - \mb b_v^T \mb x  - \mb b_h^T \mb h  % joint, not conditional!
  ,
\end{equation}
with associated probability distribution $p_{R}\left( \mb x, \mb h \right) = \frac{1}{Z_{R}}\exp\left[  -E_{\mathrm{R}}(\mb x, \mb h) \right]$. Since the there are no connections between hidden units (hence ``restricted'' Boltzmann machine), the latent variables $\mb h$ can be analytically marginalized out of the distribution (see Appendix A) to obtain 
\begin{equation}
   p\left( \mb x \right) = \int d\mb h\, p\left( \mb x, \mb h \right) = \frac{1}{Z_{R}}\exp\left[  -E_{\mathrm{R}}(\mb x) \right]
 ,
\end{equation}
where the energy for the marginalized distribution over $\mb x$ (sometimes referred to as the free energy in machine learning literature) is
\begin{equation}
    E_{\mathrm{R}}(\mb x) =  -\sum_i \log(1 + e^{\mb w_i^T \mb  x + b_{h,i}}) - \mb b_v^T \mb x
   , \label{eq:rbm}
\end{equation}
where $\mb w_i$ are rows of the matrix $\mb W$.
The energy function for an sRBM combines the Ising model and RBM energy terms,
% again artificially flip Ising signs
\begin{equation}
    E_{\mathrm{S}}\left(\mb x, \mb h\right) =  - \mb x^T \mb J \mb x  -  \mb x^T \mb W \mb h - \mb b_v^T \mb x  - \mb b_h^T \mb h
  .
\end{equation}
As with the RBM, it is straightforward to marginalize over the hidden units for an sRBM,
\begin{align}
   p_{\mathrm{S}}\left( \mb x \right) & =  \frac{1}{Z_\mathrm{S}}\exp\left[  -E_{\mathrm{S}}(\mb x) \right] 
,\\
    E_{\mathrm{S}}(\mb  x) & =  - \mb x^T \mb J \mb x - \sum_i \log(1+e^{\mb w_i^T \mb  x + b_{h,i}}) - \mb  b_v^T \mb  x
  .
\end{align}
A hierarchical Markov Random Field based on the sRBM has previously been applied as a model for natural image patches \cite{Osindero2008}, with the parameters estimated using contrastive divergence (CD, \cite{Hinton2002}). 

Instead of CD, which is based on sampling, we train the models using Minimum Probability Flow (MPF, \cite{Sohl-Dickstein2011}), a recently developed estimation method for energy based models. MPF works by minimizing the KL divergence between the data distribution and the distribution which results from moving slightly away from the data distribution towards the model distribution.  This KL divergence will be uniquely zero in the case where the model distribution is identical to the data distribution.  While CD is a stochastic heuristic for parameter update, MPF provides a deterministic and easy to evaluate objective function. Second order gradient methods can therefore be used to speed up optimization considerably.  The MPF objective function
\begin{align}
K & = \sum_{\mb x \in \mathcal D} \sum_{\mb x' \notin \mathcal D} g\left( \mb x, \mb x' \right) \exp\left( \frac{1}{2}\left[
E(\mb x) - E(\mb x') \right] \right)
\end{align}
measures the flow of probability out of data states $\mb x$ into neighboring non-data states $\mb x'$, where the connectivity function $g\left( \mb x, \mb x' \right) \in \left\{ 0, 1 \right\}$ identifies neighboring states, and $\mathcal D$ is the list of data states.  We consider the case where the connectivity function $g\left( \mb x, \mb x' \right)$ is set to connect all states which differ by a single bit flip %, or states with all bits flipped
\begin{align}
g\left( \mb x, \mb x' \right)
 =
	\left\{\begin{array}{ccrl}
1 & & H\left( \mb x, \mb x' \right) = 1  \\
0 & & \mathrm{otherwise} & 
	\end{array}\right.
,
\end{align}
where $H\left( \mb x, \mb x' \right)$ is the Hamming distance between $\mb x$ and $\mb x'$.  See Appendix B\footnote{Code is available for download at http://github.com/Sohl-Dickstein} for a derivation of the MPF objective function and gradients for the sRBM, RBM, and Ising models. In all experiments, minimization of K was performed with the MinFunc implementation of L-BFGS~\cite{Schmidt2005}.

To prevent overfitting all models were estimated with an $L_1$ sparseness penalty %, corresponding to a Laplace prior, % JSD this isn't exactly equivalent to using a Laplace prior -- better just not to mention a probabilistic interpretation as a prior I think
on the coupling parameters. This was done by adding a term of the form $ \lambda \sum_{i,j} |J_{ij}| + \lambda \sum_{j,k} |W_{jk}| $ to the objective function, summing over the absolute values of the elements of both the visible and hidden weight matrices. The optimal sparseness $\lambda$ was chosen by cross-validating the log-likelihood on a holdout set.

% log-likelihood
Since MPF learning does not give an estimate of the partition function, we use annealed importance sampling (AIS, \cite{Neal2001}) to compute normalized probabilities. AIS is a sequential Monte Carlo method that works by gradually morphing a distribution with a known normalization constant (in our case a uniform distribution over $\mb x$) into the distribution of interest. See Appendix C for more detail. AIS applied to RBM models is described in~\cite{Salakhutdinov2008}, which also highlights the shortcomings of  previously used deterministic approximations. 

Normalizing the distribution via AIS allows us to compute the log likelihood of the model $p_{\mathrm m}$ and compare it to the likelihood gained over a baseline model.  This baseline assumes cells to be independent and characterized by their firing rate $p_{\mathrm r}(\mb x)=\prod_i ( r_i x_i + (1-r_i)(1-x_i) ) $ with rates $r_i$ for individual cells $i$. 
The independent model is easily estimated and normalized, and is commonly used as a reference for model comparison. The excess log likelihood over this baseline is defined in terms of an expectation per unit time as $\mathcal L = \frac{1}{N\tau} \sum_{\mb x \in \mathcal D}  \left[ \log_2 p_{\mathrm m}(\mb x) - \log_2 p_{\mathrm r}(\mb x) \right]$, where $\tau$ is the width of a time bin. %JSD not sure if tau is the best choice of symbols here?
The excess log likelihood rate $\mathcal L$, measured in bits/s, is used as the basis for model comparisons.

% ----------------------------------
% Results
% ----------------------------------

\section{Results}

% general bit about data and how it was collected
We estimated Ising, RBM and sRBM models for populations of cortical cells simultaneously recorded across all cortical layers in a microcolumn. Here we present data from two animals, one with 23 single units (B4), another with 36 units (T6), as well as a multiunit recording with 28 units (B4MU). In each case the population was verified to be visually responsive and the majority of cells were orientation selective simple or complex cells.

% fig2: coupling matrix
\begin{figure*}
  \centering
  \includegraphics{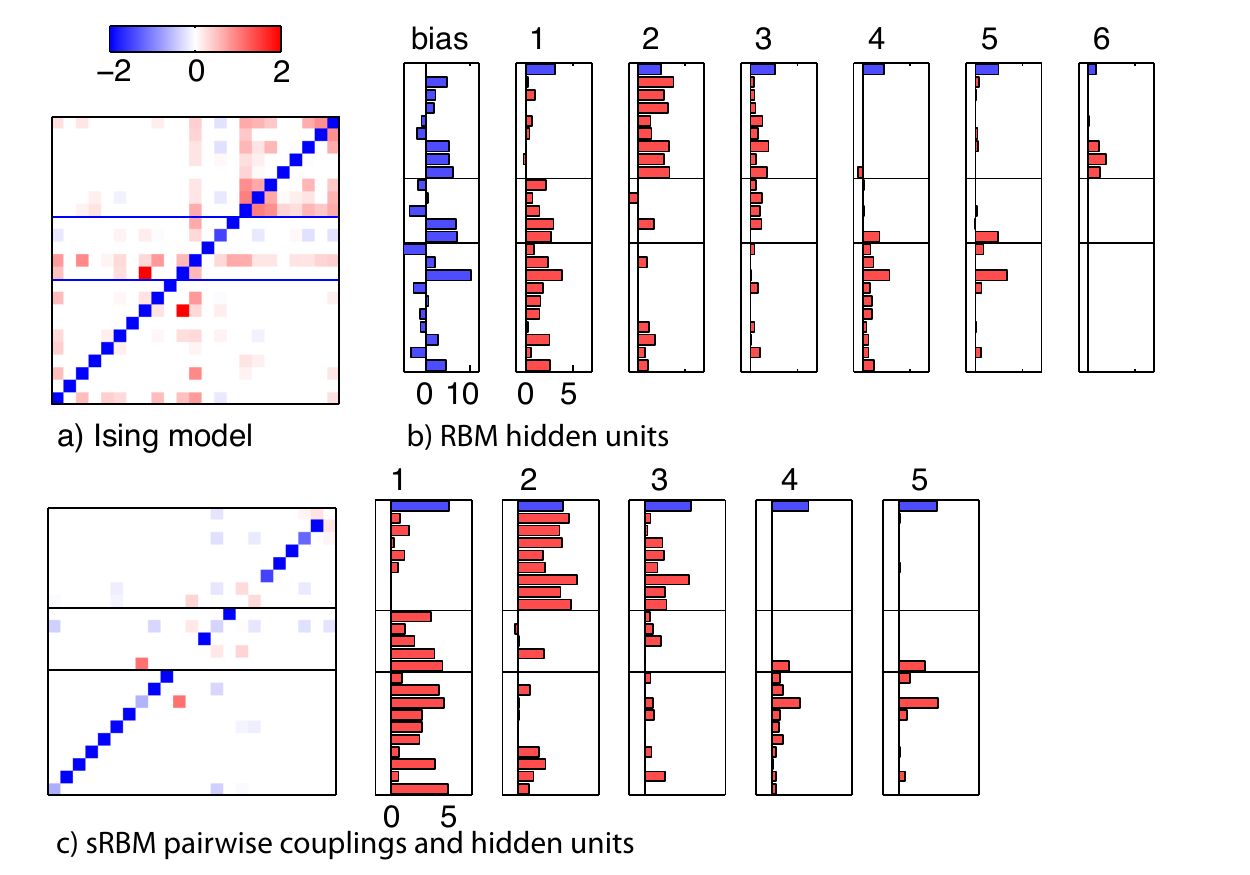} 
  \caption{ Connectivity patters of the three models estimated for a recording session with 23 spike sorted single units. The horizontal lines indicate approximate boundaries between cortical layers II/III, layer IV and layers V/VI. \textbf{(a)}  Ising model with the bias terms on the diagonal. The model has many small coupling terms that encode positive correlations. \textbf{(b)} The RBM coupling weights are plotted as a bar chart for each hidden unit, ordered by activity from left to right. The first bar chart is the bias for all the visible units, and the extra bar at the top of each plot corresponds to the bias of that hidden unit. Blue bars indicate a sign flip (the bias terms are predominantly negative, but plotted with reversed sign for easier comparison with the remaining terms). 
  % JSD "Blue bars indicate a negative sign"  confused by this.  if blue means negative, then what does being left or right of the vertical line mean?  I think blue just means bias unit?  I would also label more numbers on the horizontal axes of these plots -- at least 0, and one other number
  \textbf{(c)} The sRBM weights are shown in the same way, with the pairwise couplings on the left and hidden units on the right. The pairwise connections are qualitatively very different from those of the Ising model, as most of the structure is better captured by hidden units. }
  \label{fig:coupling} 
\end{figure*}

% fig3: logL 
\begin{figure*}
  \centering
  \includegraphics{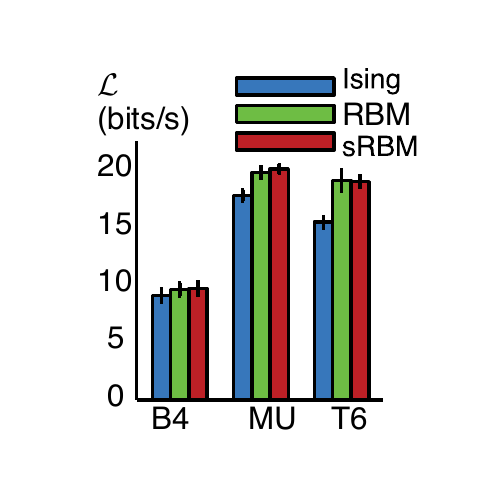} % alt contains the new model size comparison  
  \caption{  \textbf{(a)}  Model comparison using likelihood gain over the independent (firing rate) model. Likelihoods are normalized to bits/s, but not for model size, explaining the large difference between data sets of different size: 23 cells for session B4, 28 cells for MU, and 36 cells for T6. B4 and T6 are spike sorted, MU is a multiunit dataset. All three models outperform the independent model by 10-20 bits/s. The higher order models with hidden units give a further improvement of about 2 bit/s over the Ising model.  }
  \label{fig:likelihood_s} 
\end{figure*}

%isingf
The estimated model parameters for the three different types of models (Ising, RBM and sRBM) are shown in Fig.~\ref{fig:coupling}. The sparseness penalty, chosen to optimize likelihood on a validation dataset, % JSD should "test" be "validation"?
results in many of the parameters being zero. For the Ising model {\em (a)} we show the coupling as a matrix plot, with horizontal lines indicating anatomical layer boundaries. The diagonal contains the bias terms (there is no self-coupling), which are negative since all cells are off the majority of the time. The matrix has many small positive weights that encourage positive pairwise correlations by lowering the energy of connected states being active simultaneously. 

%rbm
In (b) we show the hidden units of the RBM as individual bar plots, with the bars representing connection strengths to visible units. The topmost bar corresponds to the hidden bias of the unit. Hidden units are ordered by variance. The units are highly selective in connectivity: The first unit almost exclusively connects to cells in the deep (granular and subgranular) cortical layers. The second and third unit capture correlations between cells in the superficial (supergranular) layers. The correlations are of high order, with 10 and more cells receiving input from a hidden unit. The remaining units connect fewer cells, but still tend to be location-specific. 
Unit five captures a predominantly pairwise correlation that is also visible in the Ising model coupling matrix. % JSD don't understand this.  Also should number the hidden units in the figure if going to refer to them by number.
Only six out of the total 23 hidden units have non-zero couplings and are shown.  Additional hidden units are disabled by the L1 sparseness penalty, which was chosen to maximize likelihood on the cross-validation dataset. 
The interpretation of hidden units is quite similar to the pairwise terms of the Ising model: positive coupling to a group of visible units encourages these units to become active simultaneously, as the energy of the system is lowered if both the hidden unit and the cells it connects to are active. Thus the hidden units become switched on when cells they connect to are firing (activation of hidden units not shown). 

%srbm
The sRBM combines both pairwise and hidden connections and hence is show with a pairwise coupling matrix and bar plots for hidden units. Due to the larger number of parameters, the optimal model is even more sparse. The remaining pairwise terms now predominantly encode negative interactions, whereas much of the positive coupling has been explained away by the hidden units. 
These still provide strong positive couplings within either superficial (II/III) or intermediate (IV) and deep (V/VI) layers, which explain the majority of structure in the data.
 
% jascha added this
It is noteworthy that the preferred explanation for dependencies between recorded neurons is via connections to shared hidden units, rather than direct couplings between visible units.
% JSD I would more strongly emphasize that the most conc ????????
The RBM and sRBM in this comparison were both estimated with 25 hidden units, but we show only units that did not go to zero due to the sparseness constraint. In this case, $\lambda=2 \times 10^{-3}$ was found to be optimal. Since at this level of sparseness, many of the hidden units turn off entirely, it was deemed unnecessary to further increase the number of hidden units.

% fig4: Schneidmann figure   
\begin{figure*}
  \centering
  \includegraphics{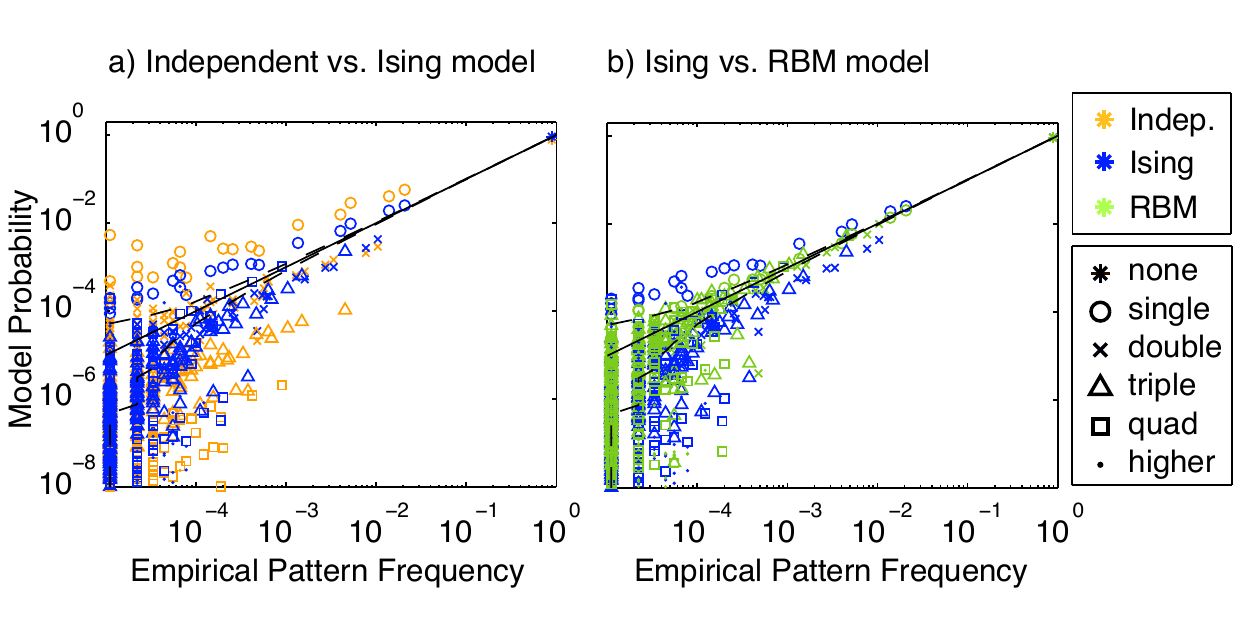} % alt contains the scatter plot 
  \caption{ Scatter plot of test data set showing empirical probabilities against model probabilities. Different models are encoded by color, the number of simultaneously spiking cells in each pattern by different symbols. \textbf{(a)} shows the independent model compared to the Ising model, \textbf{(b)} shows the Ising model compared to the RBM. The sRBM is omitted as it is very similar to the RBM. The RBM significantly outperforms simpler models.}
  \label{fig:scatter} 
\end{figure*}

% a) logL
For a quantitative comparison between models, we computed normalized likelihoods using AIS to estimate the partition function. For each model, we generated 500 samples through a chain of $10^5$ annealing steps. To ensure convergence of the chain, we use a series of chains varying the number of annealing steps and verify that the estimate of Z stabilizes to within at least 0.02 bits/bin 
(see supplemental Fig.~\ref{fig:ais} in the Appendix). For models of size 20 and smaller we furthermore computed the partition function exactly to compare against the AIS estimate. 

%We compare the likelihood of the different models as a function of the sparseness parameter $\lambda$ so that each model is optimized individually to give the maximum likelihood on a cross-validation dataset. 
Fig.~\ref{fig:likelihood_s} a) shows a comparison of excess log likelihood $\mathcal L$ over the independent firing rate model for the three different models and on all three datasets. $\mathcal L$ is computed in units of bits/s for the full population. Both higher-order models outperform the Ising model in fitting the datasets, significantly so for two datasets. Error bars are standard error on the mean, computed from 10 models with
different random subsets of the data used for learning and validation, and 
different random seeds for the parameters and the AIS sampling runs. 
% JSD this can't be right!  (unless maybe the commented out clause about random subsets should be kept?)  For instance -- the MPF objective function is convex for the Ising model case, and so there should always converge to the same parameters and same log likelihood irrespective of initialization
Each of the models was estimated for a range of sparseness parameters $\lambda=[0, 1, 2, 4, 6, 8, 10] \times 10^{-3}$ bracketing the optimal $\lambda$, and the results are shown for the optimal choice of $\lambda$ for each model. 

%While the optimal sparseness parameter $\lambda$ is largest for the Ising model, it is worth pointing out that the RBM requires significantly fewer nonzero parameters to provide a better fit to the data. In the Ising model 227/529 elements are larger than a cut-off of .001 (43\%), in the RBM 191/624 elements are above this value (31\%), and in the sRMB 285/1129 elements are active (26\%). The RBM provides a high log likelihood with the fewest parameters, showing the importance of higher order over pairwise dependencies.

% c) scatter plot
Additional insight into the relative performance of the models can be gained by comparing model probabilities to empirical probabilities for the various types of patterns. Fig.~\ref{fig:scatter} shows scatter plots of model probabilities under the different models against pattern frequencies in the data. Patterns with a single active cell, two simultaneously active cells, etc.\ are distinguished by different symbols. As expected from the positive correlations, the independent model (yellow) shown in a) has to greatly overestimate the probabilities of cells being active individually, so these patterns fall above the identity line, while all other patterns are underestimated. For comparison the Ising model is shown in the same plot (blue), and does significantly better, indicated by the points moving closer to the identity line. It still tends to fail in a similar way though, with many of the triplet patterns being underestimated as the model cannot capture triplet correlations. In b), this model is directly compared against the RBM (green). Except for very rare patterns, most points are now very close to the identity line, as the model can fully capture higher order dependencies. Hidden units describe global dependencies that greatly increase the frequency of high order patterns compared to individually active cells. 
The 5\% and 95\% confidence intervals for the counting noise expected in the empirical frequency of system states %from a binomial distribution on the model probabilities
are shown as dashed lines.  The solid line is the identity. 
% a few words on checking Z

Note that any error in estimating the partition function of the models would lead to a vertical offset of all points. Thus visually checking the alignment of the data cloud around the identity line provides an intuitive verification that there are no catastrophic errors in the estimation of the partition function. Unfortunately we cannot use this alignment (e.g. of the most frequent all zeros state) as a shortcut to compute the partition function without sampling: L$_1$ regularization tends to reduce model probabilities of the most frequent states, so this estimate of Z would systematically overestimate the likelihood of regularized models. We note, however, that for models with no regularization this estimate does indeed agree well with the AIS estimate.

\subsection{Spatiotemporal models}

% res4: Spatio-Temporal models

The same models can be used to capture spatiotemporal patterns by treating previous time steps as additional cells. Consecutive network states binned at 6.7ms are concatenated in blocks of up to 10 time steps
% JSD this was 10 time bins in section 2.1, and is ten time bins in Figure 5
, for a total network dimensionality of 100 with 10 cells. These models were cross-validated and the sparseness parameters optimized in the same way as for the instantaneous model. This allows us to learn kernels that describe the temporal structure of interactions between cells. 

\begin{figure*}
  \centering
  \includegraphics{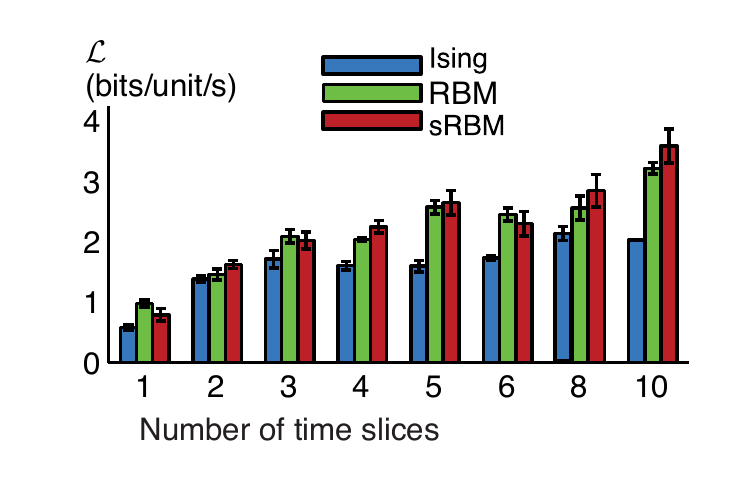} % alt contains the new model size comparison  
  % JSD -- I think y-axis should be in units \mathcal L \newline bits/(unit  sec)
  % JSD -- x-axis should be "Number of time slices"  Talk about it being for 10 units in the caption, not the axis -- it's confusing, and not the important aspect.
  \caption{Likelihood comparison as a function of model size, for spatiotemporal models with 10 cells and a varying number of concatenated time steps. The log-likelihood per neuron increases as each neuron is modeled as part of a longer time sequence. This effect holds both for Ising and higher order models, but since the Ising model cannot capture many of the relevant dependencies, the increase in likelihood saturates at about 3 timesteps.}
  \label{fig:likelihood_t} 
\end{figure*}

% b) model size
% 1. implementation details
In Fig.~\ref{fig:likelihood_t} we compare the relative performance of spatiotemporal Ising and higher order models as a function of the number of time steps included in the model. To create the datasets, we picked a subset of 10 cells with the highest firing rates from the B4 dataset (4 cells from subgranular, 2 from granuar and 4 from supergranular layers) and concatenated blocks of up to 10 subsequent data vectors. This way models of any dimensionality divisible by 10 can be estimated. The number of parameters of the RBM and Ising model were kept the same by fixing the number of hidden units in the RBM to be equal to the number of visible units, the sRBM was also estimated with a square weight matrix for the hidden layer.
% 2. describe results verbally
As before, the higher order models consistently outperform the Ising model. The likelihood per unit increases with the network size for all models, as additional information from network interactions leads to an improvement in the predictive power of the model. However, the curve levels off for the Ising model after a dimensionality of about 30 is reached, as higher order structure that is not well captured by the Ising model becomes increasingly important. 

% 3. following jascha: interpret results
A similar observation has been made in~\cite{Ganmor2011}, where Ising and higher order models for 100 retinal ganglion cells were compared to models for 10 time steps of 10 cells. It is noteworthy that temporal dependencies are similar to dependencies between different cells, in that there are strong higher order correlations not well described by pairwise couplings. These dependencies extend surprisingly far across time (at least 67ms, corresponding to the largest models estimated here) and are of such a form that including pairwise couplings to these states does not increase the likelihood of the model. 
This has implications e.g. for GLMs that are typically estimated with linear spike coupling kernels which will miss these interactions.

% JSD -- in this paragraph there is an implicit assumption that additional timesteps are equivalent to additional units at a single timestep in terms of the way the different models scale with system size.  This is interesting in terms of the retina comparison.  You should make this assumption explicit though. -- done.

% I would mostly just talk about the way in which the log likelihood and relative contribution of higher order terms scales with time, which I think is very interesting in its own right, and a more convincing story, and only mention the system size comparison as a secondary aspect.

% JSD  -- also, should point out that a GLM, at least with the linear kernel, is a pairwise model, and so that this also shows how much of the temporal structure is inaccessible to linear kernel GLMs, which is a kind of neat observation

\begin{figure*}[t] 
  \centering
  \includegraphics{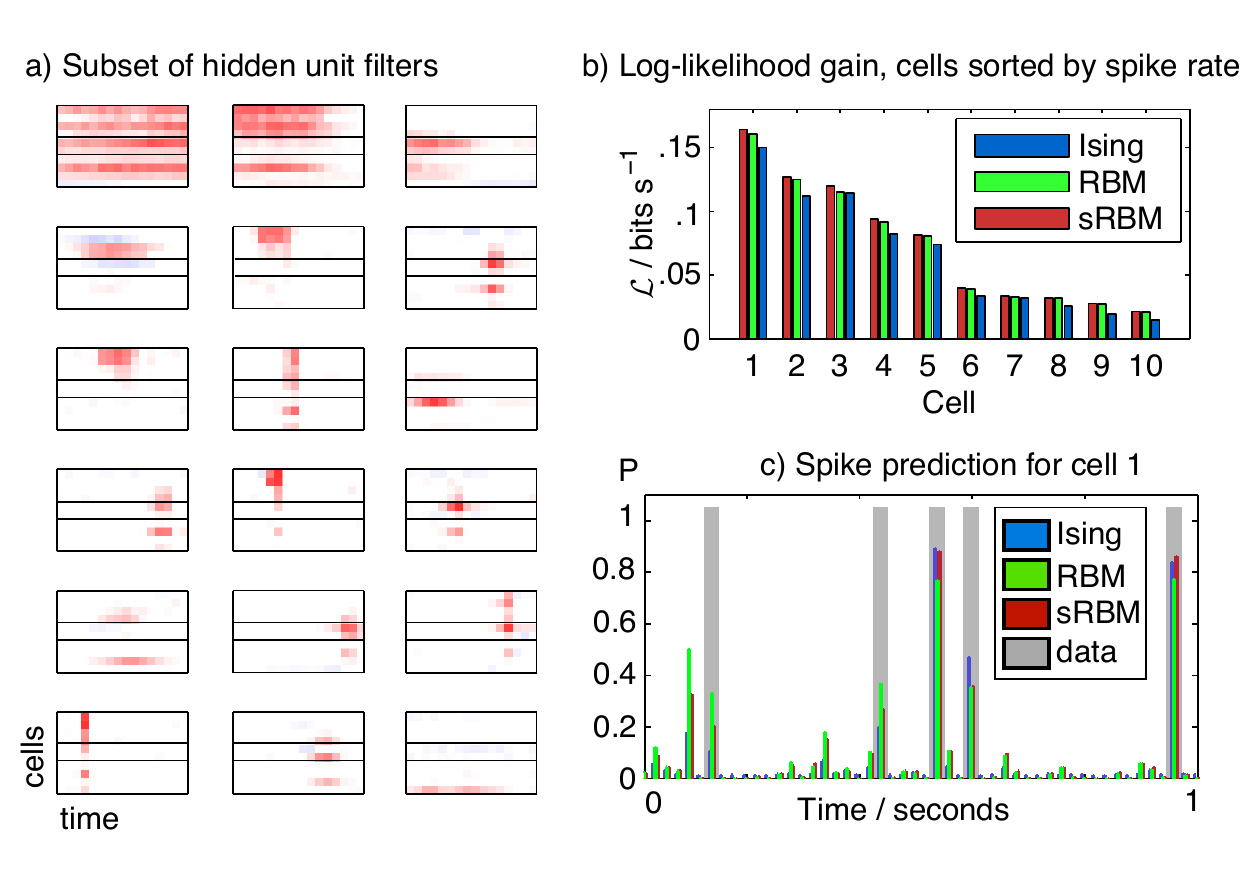} 
  \caption{ \textbf{(a)}  Hidden units of spatiotemporal sRBM model. For each hidden unit, the horizontal axis is time and the vertical axis cells with horizontal bars separating the subgranular, granuar and supergranular cortical layers. 
  % JSD are the 10 cells ordered by cortical layer?  if yes, could the layer boundaries be drawn onto this plot?  it might show some cool spatiotemporal functional connectivity structure.
  \textbf{(b)} Log-likelihood gain for one cell conditioned on the network state for all three models. 
  The remaining population carries more information about cells with higher firing rates.
%Cells are ordered by firing rate, cells with higher rates carry more information. 
%Depending on the firing rate of the cell, more additional information can be encoded by the higher order models.  
\textbf{(c)} Spike prediction from network history. For one of the cells, we show 1s of predicted activity given the history of the network state. In each case when a spike occurs in the data, there is an elevated probability under the models.}
  \label{fig:st} 
\end{figure*}

To predict spiking based on the network history, we can compute the conditional distribution of single units given the state of the rest of the network. This is illustrated for a network with 15 time steps for a dimensionality of 150.
Fig.~\ref{fig:st} a) shows the learned weights of 18 randomly chosen hidden units for a spatiotemporal RBM model with 150 hidden units. Each subplot corresponds to one hidden unit, which connects to 10 neurons (vertical axis) across 15 time steps
% JSD 15 vs 10
 or 100ms (horizontal axis). Some units specialize in spatial coupling across different cells at a constant time lag. The remaining units describe smooth, long-range temporal dependencies, typically for small groups of cells. Both of these subpopulations capture higher order structure connecting many neurons that cannot be approximated with pairwise couplings.  %For instance, it is difficult for a pairwise model to assign high probability to a group of neurons activating simultaneously without also assigning high probability to the activation of subgroups. The temporal structure occurs at time scales of 20ms to 50ms, falling into the beta and gamma frequency range. 

By conditioning the probability of one cell at one time bin on the state of the remaining network, we can compute how much information about a cell is captured by the model over a naive prediction based on the firing rate of the cell. This conditional likelihood for each cell is plotted in \ref{fig:st} b) in a similar way to excess log likelihood for the entire population in Fig.~\ref{fig:likelihood_t}. While the result here reflects our previous observation that Boltzmann machines with hidden units outperform Ising models, we note that the conditional probabilities are easily normalized in closed form since they describe a one-dimensional state space. Thus we can ensure that the likelihood gain holds independent of the estimation of $Z$ and cannot be due to systematic errors in sampling from the high-dimensional models. 
Fig.~\ref{fig:st} c) provides a more intuitive look at the prediction. For 1s of data from one cell, where 5 spikes occur, we show the conditional firing probabilities for the 3 models given 100ms of history of itself and the other cells. %The models perform remarkably well in correctly predicting a high spiking probability in the time bins when the cell is actually firing.  
Qualitatively, the models perform well in predicting spiking probabilities, suggesting it might compare favorably to prediction based on GLM-type or Ising models~\cite{Truccolo2009}. 
% JSD could this last sentence be quantified or referenced somehow?

% ----------------------------------
% Conclusion
% ----------------------------------

\section{Discussion}

% two problems
We explored the utility of Boltzmann machines with hidden units as models for neural population data, arguing that it provides a better model for cortical data than Ising models and previous extensions. While there has been a resurgence of interest in these maximum entropy models for describing neural data, progress has been hampered mainly by two problems:
% partition function
Estimation of energy based models is difficult since these models cannot be normalized in closed form. Evaluating the likelihood of the model thus requires approximations or a numerical integral over the exponential number of states of the model, making maximum likelihood estimation computationally intractable. Therefore even the pairwise Ising model is hard to estimate in general, and various approximations have been used to overcome this problem. 
% higher order parameters
This is a purely computational difficulty, but there is a more fundamental issue with generalizing models to include higher order dependencies. While this does not by itself make the model estimation any more difficult, in general the number of model parameters to be estimated is now also exponential in the size of the data. This can be dealt with either by cutting off dependencies at some low order, estimating only a small number of higher order coupling terms, or by imposing some specific form on the dependencies. 

We attempted to address both of these problems here. Parameter estimation was made tractable using MPF, and latent variables were shown to be an effective way of capturing high order dependencies. This addresses several shortcomings that have been identified with the Ising model. 

%[U] maybe say a few words about RBMs here then? --- Or move to discussion. ---
%In the neural network community on the other hand, modeling of complex, structured data such as images and speech is a long standing problem that has been approached with a variety of models and estimation methods. For binary data models of the Boltzmann machine family are widely used. A fully visible Boltzmann machine (i.e. one with no hidden units) corresponds to the Ising model, on the other end of the spectrum there are deep Bolzmann machines with multiple hidden layers. Steady progress on the estimation of these models has lead to algorithms that can train networks with millions of parameters and thosands of input units (cite google). Thus the model we consider here is fairly modest in comparison with state of the art networks, both in architectural complexity and model size. 

% [U] Give it a neuroscience spin ?

% shortcoming 1: direct coupling
As Macke argues in \cite{Macke2011}, models with direct (pairwise) couplings are not well suited to model data recorded from cortical networks. Since only a tiny fraction of the neurons making up the circuit are recorded, most input is likely to be common input to many of the recorded cells rather than direct synapses between them. 
Whiles his work compares Generalized Linear Models (GLMs) such as models of the retina~\cite{Pillow2008} and for LGN~\cite{Butts2011} to linear dynamical systems (LDS) models, the argument applies equally for the models presented here. 
%
% Similar to our RBM the LDS has a low-dimensional latent state rather than direct coupling capturing the structure of the data. While it is hard to directly compare this model of temporal dynamics to the maximum entropy models that are primarily designed to capture instantaneous correlations, the LDS model requires EM for the model estimation with a Laplace approximation to the posterior, whereas the RBM can be fit without the use of approximations.  % JSD -- MPF is an approximation to the maximum likelihood objective -- so I don't think it's fair to say we don't use approximations.  I think we can just leave this sentence off though.  I don't think there's need for a specific compare and contrast -- but if you want to do one, it's probably enough just to note that the GLM and LDS models are causal temporal models, and ours is not.

% shortcoming 2: number of parameters
Another shortcoming of the Ising model and some previous extensions is that the number of parameters to be estimated does not scale favorably with the dimensionality of the network. The number of pairwise coupling terms in GLM and Ising models scales with the square of the number of neurons, so with the amounts of data typically collected in electrophysiological experiments it is only possible to identify the parameters for small networks with a few tens of cells. This problem is aggravated by including higher order couplings: for example the number of third order coupling parameters scales with the cube of the data dimensionality. Therefore attempting to estimate these coupling parameters directly is a daunting task that usually requires approximations and strong regularization.

%now bring up Elad&Elad work after setting the context, haha!  Ohiorhenuan
An alternative is to focus on higher order structure in very small networks. Ohiorhenuan noted that Ising models fail to explain structure in cat visual cortex~\cite{Ohiorhenuan2010} and was able to model triplet correlations~\cite{Ohiorhenuan2011} by considering very small populations of no more than 6 neurons. However, Schneidman and Ganmor caution~\cite{Ganmor2011} that trying to model small subsets (10 cells) of a larger network to infer properties of the full network may lead to wrong conclusions and show that for retinal networks higher order correlations start to dominate only once a certain network size is reached. Therefore they address the same question as the present paper, i.e.\ how to capture $n^\mathrm{th}$ order correlations without the accompanying $d^n$ growth in the number of free parameters in a larger network. In their proposed \emph{reliable interaction model}, they exploit the sparseness of the neural firing patterns to argue that most 
%n-wise coupling terms $\beta_{i,j,k,...}x_i x_j x_k ...$ will be zero. 
higher order coupling terms will be zero.
Therefore the true distribution can be well approximated from a small number of these terms, which can be calculated using a simple recursive scheme. In practice, the main caveat is that only patterns that appear in the data many times are used to calculate the coupling terms. While the model by construction assigns correct relative probabilities to observed patterns, the probability assigned to unobserved patterns is unconstrained, and the most probable states may therefore be ones which never occur in the data.

%, again precluding model estimation for very high dimensional data. Mathematically, this heuristic comes with no consistency guarantees. The resulting energy based model does not correspond to a correctly normalized probability distribution, so likelihoods computed from the model could be systematically biased. While the model by construction assigns correct relative probabilities to observed patterns, it is in general not possible to guarantee the non-existence of spurious minima in the energy of the model leading to large deviations of the estimated from the true partition function. % (why is this a problem?) most likely the strong l0 sparse construction prevents this, though! "not satisfactory from a mathematical point of view"? 

% now use this to plug structured models. 
% ---
% DG model: 
% Highdimensional gaussian. Each 2D projection captures correlation between a pair of cells. Thresholding samples from the gaussian produces on or off spikes. Again this is some kind of hack where n^2 parameters in the full covariance reproduce the _pairwise_ correlations. 
% in one way this is a poor-mans approximation to max-ent since it's cheap to fit and sample from (but Yu only use 10 cells anyway) but the implicit higher order correlations (which are minimized in a true Ising) do a good job of capturing common input and Yu shows that it fits the data very well. 
% ---
In contrast to these models focussed on retinal data however, in visual cortex higher order correlations play an important role even in small networks. 
Yu et al.~\cite{Yu2008,Yu2011} show that over the scale of adjacent cortical columns of anesthetized cat visual cortex, small subnetworks of 10 cells are better characterized with a dichotomized Gaussian model than the pairwise maximum entropy distribution. While the dichotomized Gaussian~\cite{Macke2009} is estimated only from pairwise statistics, it carries higher order correlations that can be interpreted as common Gaussian inputs~\cite{Macke2011b}. However these correlations are implicit in the structure of the model and not directly estimated from the data as with the RBM, so it is not clear that the model would perform as well on different datasets. 

% GLM section taken from intro. 
Finally, GLMs~\cite{Pillow2008} can be used to model each cell conditioned on the rest of the population. While mostly used for stimulus response models including stimulus terms, they are easily extended with terms for cross-spike coupling, which capture interactions between cells. 
A major limitation of GLMs is that current implementations can only be estimated efficiently if they are linear in the stimulus features and network coupling terms, so they are not easily generalized to higher order interactions. 
Two approaches have been used to overcome this limitation for stimulus terms. The GLM can be extended with additional nonlinearities, preserving convexity on subproblems~\cite{Butts2011}. Alternatively, the stimulus terms can be packaged into nonlinear features which are computed in preprocessing and usually come with the penalty of a large increase in the dimensionality of the problem~\cite{Gerwinn2010}. However, we are not aware of any work applying either of these ideas to spike history rather than stimulus terms. 
Another noteworthy drawback of GLMs is that instantaneous coupling terms cannot be included as this causes probabilities to diverge~\cite{Macke2011}, so instantaneous correlations cannot be modeled and have to be approximated using very fine temporal discretization.  % JSD of course our model runs with a pretty coarse temporal bin too

% And conclude
In conclusion, the RBM provides a parsimonious model for higher order dependencies in neural population data. Without explicitly enumerating a potentially exponential number of coupling terms or being constrained by only measurements of pairwise correlations, it provides a low-dimensional, physiologically interpretable model that can be easily estimated for populations of 100 and more cells. 

% and a bit of speculation
The connectivity patterns the RBM learns from cells simultaneously recorded from all cortical layers are spatially localized, showing that small neural assemblies within cortical layers are strongly coupled. This suggests a shared computation these local networks perform on common input, while cells across different cortical layers participate in distinct computations and have much less coupled activity. This novel observation is made possible by the RBM: because each of the hidden units responds to (and therefore learns on) a large number of recorded patterns, it can capture dependencies that are too weak to extract with previous models. In particular, the connectivity patterns discovered by the RBM and sRBM are by no means obvious from the covariance of the data or by inspecting the coupling matrix of the Ising model. This approach, combining a straightforward estimation procedure and a powerful model, can be extended from polytrode recordings to capture physiologically meaningful connectivity patterns in other types of multi-electrode data.

\subsection*{Acknowledgments}
BAO, CMG and UK were supported by National Eye Institute grant \#EY019965.

\bibliographystyle{unsrt} % plain.bst or apalike
\bibliography{/Users/urs/Dropbox/mendeley} % mendeley 

% ----------------------------------
% Appendices (after references
% ----------------------------------

\appendix

% ----------------------------------
% Integral for RBM
% ----------------------------------
\section{Marginalizing over hidden units}
In this section we review how the joint probability for visible and hidden units in the RBM can be integrated in closed form to obtain the marginal distribution for the visible units. First we rewrite the sum over hidden variables as a product
\begin{align}
    p(\mb x, \mb h) &= \frac{1}{Z} \exp \left( \sum_{i,j} x_i h_j w_{ij} + \sum_i x_i a_i + \sum_j h_j b_j \right)  \\
          &= \frac{1}{Z} e^{\sum_i x_i a_i} \prod_{j=1}^M \exp \left( \sum_i x_i h_j w_{ij} + h_j b_j \right)
\end{align}
and integrate $p(\mb x) = \sum_{\{h\}} p(\mb x, \mb h)$, where the sum is over all configurations of the hidden units. 
For the purpose of this derivation, we move the first term outside the product
\begin{equation}
    p(\mb x) = \frac{1}{Z} e^{\sum_i x_i a_i}
    	\sum_{\{h\}}  
               \exp \left( \sum_i x_i h_1 w_{i1} + h_1 b_1 \right) 
\prod_{j=2}^M  \exp \left( \sum_i x_i h_j w_{ij} + h_j b_j \right).
\end{equation}
We can write the sum over all hidden states as nested sums over every state for each hidden unit
% -- insert missing step here
\begin{align}
    p(\mb x) =  \frac{1}{Z} e^{\sum_i x_i a_i}
     \sum_{h_1 \in \{0,1\}} ... \sum_{h_m \in \{0,1\}} 
               \exp \left( \sum_i x_i h_1 w_{i1} + h_1 b_1 \right) \\
\prod_{j=2}^M  \exp \left( \sum_i x_i h_j w_{ij} + h_j b_j \right)
.
\end{align}
% --
Noting that the term we have singled out appears only in one of the sums, we rearrange to isolate the sum,
\begin{align}
    p(\mb x) = & \frac{1}{Z} e^{\sum_i x_i a_i}  \sum_{h_1 \in \{0,1\}} \exp \left( \sum_i x_i h_1 w_{i1} + h_1 b_1 \right) \\
& \sum_{h_2 \in \{0,1\}} ... \sum_{h_m \in \{0,1\}} 
\prod_{j=2}^M  \exp \left( \sum_i x_i h_j w_{ij} + h_j b_j \right).
\end{align}
Doing this for all terms we obtain a product over individual one-dimensional sums
\begin{align}
    p(\mb x) &= \frac{1}{Z} e^{\sum_i x_i a_i} \prod_{j=1}^M \sum_{h_j \in \{0,1\}}  \exp \left( \sum_i x_i h_j w_{ij} + h_j b_j \right) \\
    	&= \frac{1}{Z} e^{\sum_i x_i a_i} \prod_{j=1}^M \left[ 1 + \exp \left( \sum_i x_i w_{ij} +  b_j \right) \right] \\
    	&= \frac{1	}{Z}
	\exp \left(  \sum_i x_i a_i  
		\sum_{j=1}^M \log \left[ 1 + \exp \left( \sum_i x_i w_{ij} +  b_j \right) \right]
		\right)
    .
\end{align}
The marginal distribution over $\mb x$ for the RBM is now in the form of a standard energy based model, with energy function
\begin{align}
    E(\mb x) = -  \sum_i x_i a_i  
		-\sum_{j=1}^M \log \left[ 1 + \exp \left( \sum_i x_i w_{ij} +  b_j \right) \right]
    .
\end{align}
The sRBM follows the same logic, since additional terms in the energy that do not depend on the hidden units stay outside the sum over hidden configurations in the same fashion as for the visible bias.

% ----------------------------------
% Ising
% ----------------------------------
\section{MPF objective}
\subsection{Ising model}
Here we review the derivation of the MPF objective for an Ising model, where the objective function consists of terms connecting the data states to all states which differ by a single bit flip.

%-----
The general MPF objective function is given by
\begin{align}
K \left( \mb \Theta \right) & = \sum_{\mb x \in \mathcal D} \sum_{\mb x' \notin \mathcal D} g \left( \mb x, \mb x' \right) 
\exp \left( \frac{1}{2}
\left[ E(\mb x; \mb \Theta ) - E(\mb x';\mb \Theta) \right] 
\right)
,
\end{align}
where $g\left( \mb x, \mb x' \right) = g\left( \mb x', \mb x \right) \in \left\{ 0, 1 \right\}$ is the connectivity function, $E(\mb x; \mb \Theta )$ is an energy function parameterized by $\mb \Theta$, and $\mathcal D$ is the list of data states.  We consider the case where the connectivity function $g\left( \mb x, \mb x' \right)$ is set to connect all states which differ by a single bit flip,
\begin{align}
g\left( \mb x, \mb x' \right)
 =
	\left\{\begin{array}{ccrl}
1 & & \mb x \mathrm{\ and\ } \mb x' \mathrm{\ differ\ by\ a\ single\ bit\ flip,\ }  & \sum_n \left| x_n - x_n' \right| = 1  \\
0 & & \mathrm{otherwise} & 
	\end{array}\right.
\end{align}
The MPF objective function in this case is
\begin{align}
K\left( \mb \Theta \right) = \sum_{\mb x \in \mathcal D} \sum_{n=1}^N \exp\left( \frac{1}{2}\left[
E(\mb x; \mb \Theta) - E(\mb x + {\mb d}(\mb x, n); \mb \Theta) \right] \right)
\label{eq:K}
\end{align}
where the sum over $n$ is a sum over all data dimensions, and the bit flipping function ${\mb d}(\mb x, n) \in \left\{ -1, 0, 1 \right\}^N$ is
\begin{align}
{\mb d}(\mb x, n)_i =
	\left\{\begin{array}{ccc}
0 & & i \neq n \\
-(2 x_i - 1) & & i = n
	\end{array}\right.
\end{align}
% -----

For the Ising model, the energy function is
\begin{align}
E = \mb x^T \mb J \mb x
\end{align}
where $\mb x \in \left\{ 0, 1 \right\}^N$, $\mb J \in \mathcal R^{N\times N}$, and $\mb J$ is symmetric ($\mb J = \mb J^T$). The bias terms have been absorbed into the diagonal of the matrix $\mb J$ which is possible since $x^2=x$ holds for binary $\mb x$.

Substituting this energy into the MPF objective function, it becomes
\begin{align}
K & = \sum_{\mb x \in \mathcal D} \sum_n \exp\left( \frac{1}{2}\left[
\mb x^T \mb J \mb x
- (\mb x + {\mb d}(\mb x, n))^T \mb J (\mb x + {\mb d}(\mb x, n))
\right] \right) \\
& = \sum_{\mb x \in \mathcal D} \sum_n \exp\left( \frac{1}{2}\left[
\mb x^T \mb J \mb x
- \left(
\mb x^T \mb J \mb x
+
2 \mb x^T \mb J {\mb d}(\mb x, n)
+
{\mb d}(\mb x, n)^T \mb J {\mb d}(\mb x, n)
\right) 
\right] \right)
 \\
& = \sum_{\mb x \in \mathcal D} \sum_n \exp\left( -\frac{1}{2}\left[
2 \mb x^T \mb J {\mb d}(\mb x, n)
+
{\mb d}(\mb x, n)^T \mb J {\mb d}(\mb x, n)
\right]
\right)  \\
& = \sum_{\mb x \in \mathcal D} \sum_n \exp\left( -\frac{1}{2}\left[
2 \sum_i x_i J_{in} \left( 1 - 2 x_n  \right)
+
J_{nn}
\right]
\right)\\
& = \sum_{\mb x \in \mathcal D} \sum_n \exp\left( \left[
\left( 2 x_n - 1 \right) \sum_i x_i J_{in}
-
\frac{1}{2}J_{nn}
\right]
\right)
.
\end{align}

Assume the symmetry constraint on $\mb J$ is enforced by writing it in terms of a another possibly asymmetric matrix $\mb J' \in \mathcal R^{N\times N}$,
\begin{align}
\mb J = \frac{1}{2} \mb J' + \frac{1}{2} \mb {J'}^T
.
\end{align}
The derivative of the MPF objective function with respect to $\mb J'$ is
\begin{align}
\pd{K}{{J'}_{lm}}  & =
\frac{1}{2}\sum_{\mb x \in \mathcal D} \exp\left( \left[
\left( 2 x_m - 1 \right) \sum_i x_i {J}_{im}
-
\frac{1}{2}{J}_{mm}
\right]
\right)
	\left[
		\left( 2 x_m - 1 \right) x_l
		-
		\delta_{lm} \frac{1}{2}
	\right] \nonumber \\ & \qquad 
+
\frac{1}{2}\sum_{\mb x \in \mathcal D} \exp\left( \left[
\left( 2 x_l - 1 \right) \sum_i x_i {J}_{il}
-
\frac{1}{2}{J}_{ll}
\right]
\right)
	\left[
		\left( 2 x_l - 1 \right) x_m
		-
		\delta_{ml} \frac{1}{2}
	\right]
,
\end{align}
where the second term is simply the first term with indices $l$ and $m$ reversed.

%Note that both the objective function and gradient can be calculated using matrix operations (no for loops).  See the code.

% ----------------------------------
% RBM
% ----------------------------------
\subsection{RBM}
After marginalizing out the hidden units, the energy function over the visible
units for an RBM is given by:
\begin{align}
E(\mb x) &= -\sum_i \log ( 1 + e^{ -W_i \mb x } )
\end{align}
where $W_i$ is a vector of coupling parameters and $\mb x$ is the binary input
vector. Bias terms have been omitted for readability. 

As previously, we substitute into the objective function Eq. \ref{eq:K} to obtain
\begin{align}
K = \sum_{\mb x \in \mathcal D} \sum_n \exp\left( \frac{1}{2}\left[
-\sum_i \log \left( 1 + e^{ -W_i \mb x } \right)
+
\sum_i \log \left( 1 + e^{ -W_i \mb x + W_i {\mb d}(\mb x, n)  } \right)
\right] \right) .
\end{align}

Unlike for the Ising model there is no cancellation of data and non-data energy terms, so evaluating the function and derivative requires looping over all bit flips for the data set.

% ----------------------------------
% sRBM
% ----------------------------------

\subsection{sRBM}

The energy function over the visible units for an sRBM obtained by marginalizing out the conditionally independent hidden units is 
\begin{align}
E(\mb x; \mb J, \mb W ) =  \mb x^T \mb J \mb x - \sum_i \log (1+ e^{-\mb W_i^T \mb x})
\label{sRBM energy}
\end{align}
where $\mb x  \in \left\{0, 1 \right\}^N$ is the visible state, $\mb J= \mb J^T \in \mathcal R^{N\times N}$ is a symmetric coupling matrix, and $\mb W \in \mathcal R^{M \times N}$ is a weight matrix to $M$ hidden units.  Equation \ref{sRBM energy} consists of a term capturing connections between visible units (an Ising model), and a term capturing connections to hidden units (an RBM).

The MPF objective we use again consists of energy differences between data and non-data states differing by a single bit. For the RBM this energy difference with the $n^{th}$ bit flipped is 
\begin{align}
dE_{n}^\mathrm{R} =& -\sum_i \left[ \log(1+ e^{-\mb w_i^T \mb x}) - \log( 1+ e^{-\mb  w_i^T (\mb x + d(\mb x,n)  ) } ) \right] \\
 =& -\sum_i \left[ \log(1+e^{z_i}) - \log(e^{z_i}+e^{w_{in}b_n}) \right]
\end{align}
where for notational simplicity we have defined $z_i= \mb w_i^T \mb  x$ and $b=2 \mb  x-1$. The energy difference contributed by connections between visible units (the Ising model) is 
\begin{align}
dE_{n}^\mathrm{I} = 2 b_n y_n-\frac{1}{2} J_{nn}
\end{align}
where we define the shorthand $\mb y=\mb J \mb x$ for simplicity. The total objective function is then given by a sum over samples and bit flips as
\begin{align}
K = \sum_{\mb x \in \mathcal D}  \sum_{n} \exp \left[ \frac{1}{2} (dE_{n}^\mathrm{I}+dE_{n}^\mathrm{R}) \right]
\end{align}
To compute the gradient of this objective w.r.t.\ the parameters $W$ and $J$ we not that 
\begin{align}
\pd{K}{J} =&  \sum_{\mb x \in \mathcal D} \sum_{n} K_{n}  \pd{}{J} dE_{n}^\mathrm{I} \\
\pd{K}{W} =&  \sum_{\mb x \in \mathcal D} \sum_{n} K_{n}  \pd{}{W} dE_{n}^\mathrm{R} 
\end{align}
these terms are computed as 
\begin{align} % ising term taken straight from the code, may need cleanup
\pd{}{J} dE_{n}^\mathrm{I} = \pd{}{J} \left(2 b_ny_n-\frac{1}{2} J_{nn} \right) 
=  2 b_n x_n  -  \frac{1}{2} 
\end{align}
for the pairwise terms, and 
\begin{align}
\pd{}{W_{ab}} dE_{n}^\mathrm{R} =& - \pd{}{W_{ab}} \sum_i \left[ \log(1+e^{z_i}) - \log(e^{z_i}+e^{w_{in}b_n}) \right] \\
=& \frac{1}{2} \sum_n \frac{e^{z_a}}{1+e^{z_a}} x_b \\ % A: outer product
+& \frac{1}{2} \sum_n \frac{ e^{z_a} }{ e^{z_a} + e^{w_{an} b_n} } x_b \\ % B1 add up terms
+& \frac{1}{2} \sum_j \frac{1}{ 1 + e^{z_a-w_{ab}b_b} } b_b %B2 one vector per loop iteration
\end{align}
for the higher order terms.

\section{Annealed importance sampling}

\begin{figure*}
  \centering
  \includegraphics{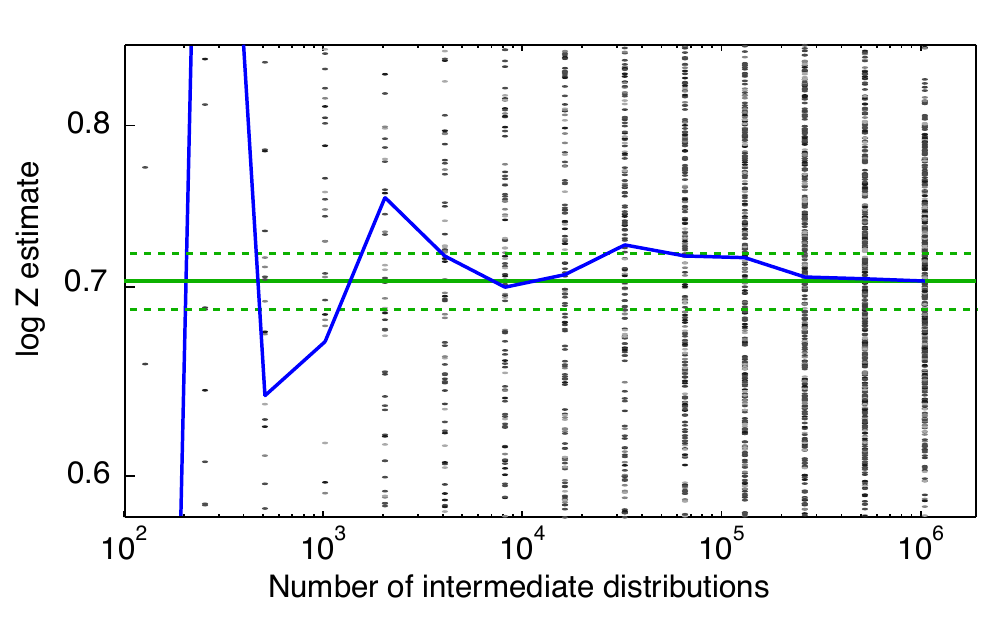} % alt contains the scatter plot 
  \caption{ Monitoring the convergence of the AIS estimate for the partition function, here for a 20-dimensional Ising model. Each entry on the horizontal axis corresponds to an annealing chain with a different number of steps. Points correspond to the 500 individual samples, the blue line is the $\log_2$ of the average from the samples. The solid green line is the true value of the partition function computed numerically by summing over the $2^{20}$ states. The dashed lines correspond to our convergence criterion of 0.02 deviation from the true partition function.}
  \label{fig:ais}
\end{figure*}

Estimating the normalization constant, also referred to as partition function, of an energy based probabilistic model, remains a challenging task~\cite{Salakhutdinov2008}. Many approaches to make learning the parameters of energy based models tractable, such as Contrastive Divergence, MPF, Score and Ratio Matching~\cite{Hyvarinen2006,Hyvarinen2007} do not attempt to estimate the partition function. A notable exception is Noise Contrastive Estimation~\cite{Gutmann2009} which treats the partition function as a parameter to be estimated, but has only been applied for continuous-valued data. Most commonly the partition function is estimated by sampling. 

Using importance sampling, the partition function can be estimated by 
\begin{align}
Z_p / Z_q = \left< \frac{\tilde{p}(\mb x)}{\tilde{q}(\mb x)} \right>_{q(\mb x)}
\end{align}
where $Z_q$ is the known partition function of the proposal distribution $q(\mb x)$, $Z_p$ is the partition function of interest for $p(\mb x)$ and the symbol\ \  $\tilde{}$\ \  indicates a non-normalized distribution. The angle brackets indicate a sample expectation over samples from the distribution $p(\mb x)$. However, if $q(\mb x)$ is not a good match to the target distribution, it takes a very large number of samples to get a good estimate. AIS uses an annealing process to gradually transform a simple proposal distribution, such as the uniform distribution, into the target distribution, leading to an accurate estimate of $Z$ from only a small number of samples. 

To assure convergence of the estimator, we run several annealing chains, increasing the number of steps in factors of 2 up to a size of $10^5$ steps. We check that the final estimate of $\log_2(Z)$ does not deviate more than 0.02 from the previous estimates. This criterium was chosen since $\log_2(Z)$ appears as an additive term to $\mathcal L$, and at a bin size $\tau=50 \mathrm{ms}$ an error of 1 bits / second in the final estimate of the likelihood was seen as an acceptable trade-off between estimation speed and accuracy. In Fig. \ref{fig:ais}, we show this convergence plot for a small 20-dimensional model, where the normalization constant was computed exactly. For larger models, where the partition function could not be calculated analytically, we monitored that the estimate stabilized to within this tolerance. 

% JSD - In section 3 it says "For each model, we generated 500 samples through a chain of 10^5 annealing steps."  This isn't consistent with the previous paragraph, where you say the number of intermediate distributions is increased until log(Z) stabilizes.

\end{document}